\documentclass[11pt]{article}
\usepackage[usenames]{color}
\usepackage{graphics,amsmath}
\usepackage[titletoc,title]{appendix}
\usepackage{epsfig}
\usepackage{multirow}
\usepackage{colordvi}
\newcommand{\removed}[1]{}

\newcommand{\bfm}[1]{\mbox{{\boldmath $ { #1}$}}}
\newcommand{\grad}{{\bfm \nabla}}

\newcommand{\pd}[2]{\frac{\partial{#1}}{\partial{#2}} }

\title{A One-Dimensional Tearing Mode Equation for Pedestal Stability Studies in Tokamaks.}

\author{ J.W. Connor$^{1,2}$, R.J. Hastie$^{1}$, C. Marchetto$^{3}$ and C.M. Roach$^{1}$ \\
\vspace*{5mm}  \\
\small $^{1}$ UKAEA-CCFE, Culham Science Centre, Abingdon, Oxon, OX14 3DB, UK \\
\small $^{2}$ Imperial College of Science, Technology and Medicine, London, SW7 2BZ, UK \\
\small $^{3}$ Istituto di Fisica del Plasma, CNR, Via Roberto Cozzi, 53, 20125 Milano, Italy
\normalsize}

\begin{document}

\maketitle

\begin{abstract}
 Starting from expressions in Connor {\em et al.} (1988) \cite{C2H3M}, we derive a one-dimensional tearing equation similar to the 
 approximate equation obtained by Hegna and Callen (1984) \cite{HC} and by Nishimura {\em et al.} (1998) \cite{NCH}, but for more realistic toroidal equilibria. 
 The intention is to use this approximation to explore the role of steep profiles, bootstrap currents and 
 strong shaping in the vicinity of a separatrix, on 
 the stability of tearing modes which are resonant in the H-mode pedestal region of finite aspect ratio, shaped cross-section  tokamaks, e.g. JET. 
 We discuss how this one-dimensional model for tearing modes, which assumes a single poloidal harmonic for the perturbed poloidal flux, compares with a model 
 that includes poloidal coupling by Fitzpatrick {\em et al.} (1993) \cite{FITZPATRICK_NF1993}.
 \end{abstract}

\section{  Introduction}

 Edge Localised Modes (ELMs) are a ubiquitous feature of H-mode tokamak plasmas with important consequences for confinement and for 
 transient heat loads on divertor target plates. Most theoretical models appeal to ideal magnetohydrodynamic (MHD) ballooning and peeling 
 modes \cite{HCHW, CHWM, Wilson99, PhilS}
 as the trigger for ELMs. While this may well be the case for larger Type I ELMs, the smaller Type III may involve 
 resistive ballooning modes \cite{CONNOR_PPCF1998}. Furthermore it is unclear whether ideal peeling modes are ever unstable due to the presence of a separatrix 
 in divertor tokamaks \cite{Guido, WEBSTER_PRL2009}
 or can lead to the required destruction of magnetic surfaces seen in resistive MHD simulations, e.g. \cite{ASDEX}. 
 However, concerning the need for resistivity, one should mention a model for ELMs in which an unstable ideal peeling mode does play a part, 
 triggering a Taylor relaxation in the edge plasma, thus involving reconnection. The relaxation region grows in size until the ideal mode becomes 
 stable \cite{GHH}. An alternative 
 possible explanation is that ELMs might be triggered by tearing modes being driven unstable by the large bootstrap current density 
that results from the pressure gradients in the H-mode pedestal.

The theory of tearing modes utilises asymptotic matching techniques \cite{FURTH_PF1963}. 
Thus solutions of the resistive 
equations (or those corresponding to more complex plasma models e.g. 
\cite{ANTONSEN_PLA1981,DRAKE_PF1983,PEGORARO_PPCF1986,CKH,PORCELLI_PF1987,FITZPATRICK_POP1989,CONNOR_PPCF2012}) that pertain near 
resonant surfaces, $m=nq(\rho_s)$, are matched to solutions of ideal MHD equations that describe the regions away from the resonance to obtain a dispersion 
relation determining their stability.  Here $m$ and $n$ are poloidal and toroidal mode 
numbers of the perturbation, $q(\rho)$ is the safety factor, $\rho$ is a flux surface label with dimensons of length, and $\rho_s$ is the resonance position. 
This matching procedure involves obtaining the asymptotic forms of the ideal MHD solutions as $\rho \to \rho_s$ from both left and right, and the matching is 
characterised by a quantity $\Delta^{\prime}$. Stability of a mode is determined by comparing $\Delta^{\prime}$ with $\Delta^{\prime}_{\rm crit}$, 
a parameter that is determined from the solution of the equation describing the narrow layer around the resonance. The quantity $\Delta^{\prime}_{\rm crit}$ is 
usually a large positive number \cite{GGJ,DRAKE_PF1983,CKH}, but physics close to the resonance can make $\Delta^{\prime}_{\rm crit}$ negative: e.g. when 
microtearing modes are unstable, as has been reported for the region around the H-mode pedestal in MAST \cite{DICKINSON_PRL2012} and JET \cite{HATCH_NF2016} plasmas.
 
The linear theory of  tearing instability in toroidal geometry \cite{C2H3M} is a complex problem, raising issues associated with the coupling of different 
poloidal harmonics and with the decoupling of resonances at different rational surfaces due to differing diamagnetic frequencies at such surfaces.  
\cite{HC} proposed a simple approximation that the perturbed poloidal flux has a single poloidal harmonic, of admittedly uncertain accuracy, to obtain a 
master equation for tearing instability, with similar one-dimensional 
(which in future we abbreviate to 1-D) character to that holding in a straight cylinder. This equation was derived for equilibria with weakly shaped poloidal 
cross-section, and under the additional assumptions of large aspect ratio, low $\beta$ (where $\beta$ is the ratio of plasma pressure, $p$, to the magnetic field 
energy density, $\beta=2 \mu_0 p/B^2$), and with the toroidal magnetic field greatly exceeding the poloidal field:
 \begin{equation}
 \frac{I}{q}\frac{d }{d \psi}\frac{q}{I} \langle g^{\psi \psi} \rangle \frac{d \tilde{A}}{d \psi}-
 \left[ m^2\langle g^{\theta \theta}\rangle +\frac{m}{m-nq}I\langle \sigma\rangle^\prime 
 +\frac{m^2}{(m-nq)^2}\mu_0 I p^\prime\langle J \rangle^\prime \right] \tilde{A}=0,
 \end{equation}
 where $\tilde{A}$ and $\psi$ are respectively the perturbed and equilibrium poloidal flux, 
 the magnetic field is ${\bf B}=I \grad \phi +\grad \phi \times \grad \psi$, $I$ is the toroidal field function $I=RB_{\phi}$, 
 $\phi$ is the toroidal coordinate, $\sigma=\mu_0 j_\parallel /B$,  
 with the parallel current density $j_{\parallel}=\bfm{j.B}/B$, $q$ is the safety factor,
 $^{\prime}$ denotes the radial gradient with respect to $\psi$,
 $\langle Y \rangle$ is the flux surface average of $Y$ for any quantity $Y(\psi,\theta)$,
 \begin{equation}
 \langle Y \rangle=\frac{1}{2 \pi}\oint Y d \theta,
 \end{equation}
and the metric elements are $g^{\psi \psi}=|\nabla \psi|^2$ and $g^{\theta \theta}=|\nabla \theta |^2$.  
The $\theta$ coordinate is a straight field line poloidal angle and $J=(\nabla \psi \times \nabla \theta. \nabla \phi)^{-1}=R^2 q/I$ is the Jacobian.
\cite{NCH}  presented numerical solutions of a similar equation, for a family of equilibrium profiles resembling those studied 
previously by \cite{FRS}  in cylindrical geometry.
 
To assist the tearing mode stability analysis of the H-mode pedestal, in this paper we develop a 1-D ideal MHD equation 
for application to realistic, fully toroidal tokamak equilibria at high $\beta$, thus generalising the earlier seminal works by \cite{HC} and \cite{NCH}. 
This contrasts with alternative approximate treatments described in \cite{FITZPATRICK_NF1993}, where the effect of poloidal mode 
coupling was calculated for toroidal equilibria of large aspect ratio, low $\beta$, and weak shaping, and approximate solutions with  
seven poloidal harmonics were used to obtain $\Delta^{\prime}$.
 
\section{  A 1-D Tearing Mode Equation.}
 We start from eqns.(A5) and (A6) of \cite{C2H3M}, which respectively govern the radial component of the displacement ($\bfm{\xi}$), and 
 the perturbed toroidal magnetic field: these quantities manifest themselves in Connor {\em et al}'s variables $y=R_0 f \bfm{\xi}.\grad \rho$ and 
 $z=R^2\delta {\bf B.\nabla}\phi / B_0$.  Here the equilibrium magnetic field is written as 
 ${\bf B}=R_0B_0\left[ g{\bf  \grad} \phi +f{\bf  \grad} \phi \times \grad \rho \right]$, where $\rho$ is a flux surface label with dimension of length,
 $B_0 g(\rho) R_0/R$ is the full toroidal magnetic field, $B_0$ is the vacuum toroidal field at the major radius, $R_0$, of the magnetic axis,  
 and $q=\frac{\rho}{R_0} \frac{g}{f}$. 
 The variable $y$, which is related to the perturbed poloidal flux, now denoted $\psi$, by $y=\psi/(m-nq)$ 
 \footnote{The behaviour of $\psi$ near the resonant surface at 
 $m=nq(\rho)$, is a combination of large and small solutions \cite{GGJ}, and it is this combination that must be matched to the solution in the inner resonant layer.}, 
 is assumed to contain only a single poloidal 
 harmonic, $e^{i m \theta}$, where $\theta$ is the poloidal angle in straight field line coordinates. 
 These equations can be used to generate the 1-D ideal MHD equation for $\psi$.

 Then eqns.(A5) and (A6) of \cite{C2H3M} take the form
 \begin{eqnarray}
 i\frac{d \psi}{d \rho}e^{i m \theta}&=& -\frac{\partial}{\partial \theta}
 \left[ \psi e^{i m \theta} \left(  i T + \frac{U}{m-nq}\right)\right]+
 \left(S z+\frac{\partial}{\partial \theta} Q \frac{\partial z}{\partial \theta}\right) \label{eq:1dpsi}\\ 
 \left(\frac{\partial}{\partial \theta}-inq\right)\frac{\partial z}{\partial \rho}&=&\psi e^{i m \theta} \left[ i W+\frac{X}{m-nq}+(m-nq)V\right]-i \psi e^{i m \theta}  \frac{\partial V}{\partial \theta}\nonumber \\
 &+&U\frac{\partial z}{\partial \theta}  -\left(\frac{\partial}{\partial \theta}-i n q\right)\left[ T^* \frac{\partial z}{\partial \theta}\right] \label{eq:1dz},
 \end{eqnarray}
 where the equilibrium quantities, $Q, S, T, U, V, W, X$ are defined in eqn.(A7) of \cite{C2H3M},
\begin{eqnarray}
  S&=&in\rho/R_0, \label{eq:S}\\
   Q&=&-i\frac{ R_0}{n\rho}  \frac{1}{|\nabla \rho |^2}, \label{eq:Q}\\
   T&=& \frac{\nabla \theta.\nabla \rho}{|\nabla \rho |^2}+i \frac{R_0 g'}{n\rho f} \frac{1}{|\nabla \rho |^2}, \label{eq:T}\\
   U&=&\frac{\mu_0 p'}{B_0^2 f^2} \frac{R^2}{R_0^2 |\nabla \rho | ^2 }, \label{eq:U}\\
   V&=&i \frac{n}{\rho R_0} \frac{R_0^2}{R^2 |\nabla \rho |^2}-i\frac{R_0}{\rho n}(\frac{g'}{f})^2  \frac{1}{| \nabla \rho |^2} \label{eq:V}\\
   W&=&\frac{2 \mu_0 p' g'}{B_0^2 f^3} \frac{R^2}{R_0^2 | \nabla \rho |^2} -\frac{d}{d\rho}\frac{g'}{f} \label{eq:W}\\
   X&=&i\frac{n \mu_0 p' \rho}{B_0^2 f^2 R_0} \left[ \frac{\partial}{\partial \theta}\left(T^*\frac{R^2}{R_0^2}\right) 
   + \frac{\partial }{\partial \rho}  \frac{R^2}{R_0^2} - \frac{R^2}{R_0^2}  \left(\frac{f'}{f}-\frac{1}{\rho} \right)-\frac{\mu_0 p'}{B_0^2 f^2}  
   \frac{R^4}{R_0^4 | \nabla \rho |^2} \right] \label{eq:X}
\end{eqnarray}
with $^{\prime}$ now representing the radial derivative with respect to $\rho$.
We note here that the above expressions were derived for equilibria of arbitrary aspect ratio, cross-sectional shape and $\beta$. The method employed in the following analysis is rather general and does not assume that the second 
dependent variable, $z$, is also of single poloidal harmonic structure.
 
To simplify the analysis we neglect the term involving $S$, relative to $m^2 Q$ in eqn.(\ref{eq:1dpsi}). This is equivalent to reducing the field line 
bending  energy in a circular cylinder from the $(m^2 + k_z^2 r^2)$ of the \cite{WA.N} analysis of stability in a linear pinch, 
to $m^2$.  
In a torus this is equivalent to an assumption that $( \epsilon/q_s)^2 \ll1$, where $\epsilon$ is the local aspect ratio and $q_s=m/n$ is 
the value of the safety factor at the resonance.  Since our focus will be on tearing modes which are resonant in the pedestal region of a tokamak of 
aspect ratio around $1/3$, $q_s$  may be of order $4$ or greater,  so this approximation would appear to introduce errors of only about  $1\%$.

The required 1-D tearing equation is now obtained by solving eqn.(\ref{eq:1dpsi}) for $\partial z /\partial \theta$, inserting the result in eqn.(\ref{eq:1dz}) 
and taking the flux surface average. Thus:
  \begin{equation}
 \frac{ \partial z}{\partial \theta}= \frac{e^{i m \theta}}{Q} \left[ \frac{1}{m}\frac{d \psi}{d\rho}+ 
 \psi \left( i T +\frac{U}{m-nq}\right)\right]-\frac{K(\rho)}{Q}, \label{eq:dzdt}
  \end{equation}
where $K(\rho)$ is a flux surface dependent constant of integration to be determined by a periodicity constraint on $z(\rho,\theta)$. Thus 
\begin{equation}
  K(\rho)=\frac{1}{m}\frac{d \psi}{d\rho} \alpha_m+\psi \left[ \gamma_m+\frac{\delta_m}{m-nq}\right], \label{eq:Kdef}
\end{equation}
with
  \begin{eqnarray}
  \alpha_m&=&\frac{\langle e^{i m\theta} |\nabla \rho|^2 \rangle }{\langle |\nabla \rho|^2 \rangle }
  =\frac{\langle \cos( m \theta) |\nabla \rho|^2 \rangle }{\langle |\nabla \rho|^2 \rangle } \label{eq:alpha_m}\\
  \gamma_m&=&i\frac{\langle e^{i m\theta}T |\nabla \rho|^2 \rangle }{\langle |\nabla \rho|^2 \rangle } =-\frac{\langle \sin( m\theta)\nabla \theta.\nabla \rho \rangle }{\langle |\nabla \rho|^2 \rangle }\\
  \delta_m&=&\frac{\langle e^{i m\theta}U |\nabla \rho|^2 \rangle }{\langle |\nabla \rho|^2 \rangle }
  =\frac{\langle \cos( m\theta) R^2  \rangle }{R_0^2 \langle |\nabla \rho|^2 \rangle }\frac{\mu_0 p^\prime}{B_0^2 f^2}, \label{eq:delta_m}
  \end{eqnarray}
where the second form in eqns.(\ref{eq:alpha_m}-\ref{eq:delta_m}) applies for equilibria which are symmetric above and below the median plane.
Now, since the $m$ number for tearing modes which are resonant in the pedestal region of a tokamak is likely to be moderately large, the coefficients defined
by $\alpha_m,~\gamma_m$ and $\delta_m$ in eqns.(\ref{eq:alpha_m}-\ref{eq:delta_m}) may be very small unless there is strong shaping. 
Consequently, we can normally neglect the integration 
constant, $K(\rho)$ defined in eqn.(\ref{eq:Kdef}).  In Section 2.1 we will investigate the consequences of retaining finite $K(\rho)$.
   
Inserting the expression for $(\partial z /\partial \theta)$ (in the $K(\rho)=0$ limit) into eqn.(\ref{eq:1dz}) and 
multiplying by the factor $e^{-i m \theta}$, we take the flux surface average to obtain a 1-D tearing mode equation.  
Expressed in terms of the equilibrium quantities, $Q, T, U, V, W$ and $X$, this takes the form:
   \begin{align}
  &\frac{(m-nq)}{m^2}\frac{d}{d\rho}\left[ \left\langle \frac{1}{Q} \right\rangle \frac{d \psi}{d\rho}\right]
  +\psi  \frac{(m-nq)}{m}\frac{d}{d\rho}\left[ i\left\langle \frac{T}{Q}\right\rangle+\frac{1}{(m-nq)}\left\langle\frac{U}{Q}\right\rangle \right] \nonumber \\
  & =\psi \left[(m-nq) \langle V \rangle+i\langle W \rangle+\frac{\langle X\rangle}{(m-nq)}+(m-nq)\left\langle\frac{T~T^*}{Q}\right\rangle
     +\frac{1}{(m-nq)}\left\langle\frac{U^2}{Q}\right\rangle \right. \nonumber \\
  &\left. +i\left\langle  \frac{U(T-T^*)}{Q}\right\rangle \right], \label{eq:nearlytear}
   \end{align}
   Now, writing $1/Q=\lambda \rho |\nabla \rho|^2$, where $\lambda=i n/R_0$ and dividing through by $\lambda(m-nq)/m^2$, eqn.(\ref{eq:nearlytear}) takes the form of 
   the second order differential equation:
   \begin{eqnarray}
   &&\frac{d}{d\rho}\left[ \rho \langle|\nabla \rho |^2 \rangle \frac{d \psi}{d\rho}\right]+m\psi\frac{d}{d\rho}\left[ i \rho \langle T | \nabla
   \rho|^2 \rangle+\frac{\rho}{(m-nq)}\langle U |\nabla \rho|^2 \rangle \right] \nonumber \\
   &&~=\psi\frac{m^2}{ \lambda} \left[ \langle V \rangle +i \frac{\langle W\rangle}{(m-nq)}+\frac{\langle X\rangle}{(m-nq)^2}\right] \nonumber\\
   &&~~+\psi m^2 \rho \left[ \langle T T^* |\nabla \rho|^2 \rangle+\frac{i\langle U(T-T^*)|\nabla \rho|^2 \rangle}{(m-nq)}
   +\frac{\langle U^2 |\nabla \rho |^2 \rangle}{(m-nq)^2} \right] \label{eq:1Dtear_gen}
   \end{eqnarray}
   which is of the same structure as the equation derived by \cite{HC}, namely
   \begin{equation}
      \frac{d}{d\rho} \left[ A(\rho) \frac{d \psi}{d\rho} \right]-\left[ B(\rho) +\frac{m C(\rho)}{(m-nq)}+\frac{m^2 D(\rho)}{(m-nq)^2} \right] \psi =0 \label{eq:1Dtear},
      \end{equation}
      where, on inserting the definitions (\ref{eq:Q}-\ref{eq:X}), 
      \begin{eqnarray}
      A&=&\rho\langle |\nabla \rho|^2\rangle \label{eq:A}\\
      B&=&\frac{m^2}{\rho} \left[ \left\langle \frac{R_0^2}{R^2} \frac{1}{|\nabla \rho|^2} \right\rangle
                          +\rho^2 \left\langle \frac{|\nabla \theta.\nabla \rho|^2}{|\nabla \rho|^2}\right\rangle \right] 
                          = m^2 \rho \langle |\nabla \theta|^2\rangle \label{eq:B}      \\
      C&=&-q\frac{d}{d\rho}\left[ \frac{R_0 g^\prime}{f}+ \frac{R_0}{fg} \frac{\mu_0 p^\prime}{B_0^2} \left\langle \frac{R^2}{R_0^2}\right\rangle \right]
      \label{eq:C}\\
      D&=&\frac{\mu_0 p^\prime}{B_0^2 f^2} \left[ \rho \frac{d}{d \rho} \left\langle \frac{R^2}{R_0^2}\right\rangle
          -\left\langle \frac{R^2}{R_0^2}\right\rangle \left( \frac{\rho g^\prime}{g} \right) \right]. \label{eq:D}
      \end{eqnarray}
Some details of the derivation of eqns.(\ref{eq:A}-\ref{eq:D}) are given in Appendix A and in Appendix B we express eqn.(\ref{eq:1Dtear})  
in terms of the variables used in eqn.(26) of \cite{HC}.
        
\subsection{ Consequences of finite K}
We now return to eqns.(\ref{eq:dzdt}) and (\ref{eq:Kdef}) and construct the additional terms that will appear in the tearing equation when 
we retain terms with finite $K(\rho)$. 
After lengthy, but straightforward, further analysis, we find that each of the coefficients $A(\rho),~B(\rho),~C(\rho)$ and $D(\rho)$ is modified 
by an additional contribution, which we shall denote by a circumflex. Thus
\begin{eqnarray}
    A\rightarrow A(\rho)-\hat{A}(\rho),\nonumber \\ 
    B \rightarrow B(\rho)-\hat{B}(\rho), \nonumber \\ 
    C\rightarrow C(\rho)-\hat{C}(\rho), \nonumber\\ 
    D \rightarrow D(\rho)-\hat{D}(\rho) 
\end{eqnarray}
with
\begin{eqnarray}
     \hat{A}(\rho)&=&A(\rho) | \alpha_m |^2  \label{eq:Acor},\\
     \hat{B}(\rho)&=&m^2A(\rho) \gamma_m^2-m\frac{d}{d\rho}\left[ A(\rho) \gamma_m \alpha_m \right],\\
     \hat{C}(\rho)&=&-q\frac{d}{d\rho}\left[ \frac{\alpha_m \delta_m A(\rho)}{q}\right],\\
     \hat{D}(\rho)&=&A(\rho)\delta_m\left(\delta_m- \frac{s \alpha_m}{\rho} \right) \label{eq:Dcor} 
\end{eqnarray}
$\hat{A}$, $\hat{B}$, $\hat{C}$, and $\hat{D}$ are small in the large $m$ limit because the numerators in the definitions 
of $\alpha_m$, $\delta_m$, $\gamma_m$ (see Eqns.(\ref{eq:alpha_m})-(\ref{eq:delta_m})) must vanish both at high $m$, or with weak shaping. 
At a fixed finite $m$ these terms can, however, become more important with stronger shaping (e.g. as one approaches the separatrix). 
      
\subsection{ Comparison with Earlier Results.}    
The Hegna-Callen equation represented a significant advance on earlier work by making possible a simple 1-D tearing 
analysis of large aspect ratio toroidal equilibria with weakly shaped poloidal cross-sections.  Our derivation has not only extended the 
validity of the 1-D equation to finite aspect ratio equilibria, subject to $\left( \epsilon/q_s \right)^2 \ll 1$ , with arbitrary poloidal shaping, 
but it has also revealed 
the presence of  new terms arising from finite values of the integration constant  $K(\rho)$. 
These additional terms of eqns.(\ref{eq:Acor}-\ref{eq:Dcor}) have no counterpart in \cite{HC} or \cite{NCH}, 
but they are small unless there is strong shaping containing poloidal harmonics that couple to the mode number, $m$.

We now compare our tearing eqn.(\ref{eq:1Dtear})  with \cite{HC} and \cite{NCH}.
We begin by transforming  from the  Hegna-Callen  %[\cite{7,8}]
equilibrium variables, $I$ and $\psi$, to the $g,~f,~\rho$ variables of the present work. Thus:
\begin{eqnarray}
I &\rightarrow& R_0 B_0 g(\rho),\label{eq:I}\\
\frac{d}{d\psi}&\rightarrow &\frac{1}{\psi^\prime}\frac{d}{d\rho},\label{eq:ddpsi}\\
\psi^\prime(\rho)&\rightarrow& R_0 B_0 f(\rho)\label{eq:psi}
\end{eqnarray}
The coefficients $A,~B,~C$ and $D$ can then be identified in eqn.(26) of \cite{HC}
and compared to eqns.(\ref{eq:A}-\ref{eq:D}). 
This shows agreement in the expressions for 
$A$ and $B$, close agreement on $C$, but not for $D$. Since 
\begin{equation}
 \sigma = \frac{1}{f} \left( g^{\prime} +\frac{g \mu_0 p^{\prime}}{B^2}\right),
\end{equation}
one can indeed write $C \propto \frac{\partial \langle \sigma \rangle }{\partial \rho}$ if $B \simeq B_{\phi}$, as in \cite{HC}.
There is some similarity with the expression for $D$ that appears in eqn.(19) of \cite{NCH}, 
where special equilibria with $g=$constant were studied so that the last term in eqn.(\ref{eq:D}) is absent,
but nevertheless their $D \propto n^2q^2$ rather than $m^2$, and so it differs away from the resonance.

As noted by Hegna, Callen and Nishimura, there is an important comparison for the expression given in eqn.(\ref{eq:D}) for $D(\rho)$. 
This  is associated with the Mercier stability criterion, $D_M < 0$,
for the ideal MHD stability of a mode localised around a rational surface \cite{Mercier}.
\cite{GGJ} showed $D_M$\footnote{The quantity labelled $D_M$ here is precisely the object denoted by $D_I$ in \cite{GGJ}.}  
plays an important role in the theory of tearing mode stability in a torus. 
They found the asymptotic form of the ideal MHD solutions as $\rho \to \rho_s$ is 
\begin{equation}
\psi \sim c_0 \left|x-1 \right|^{\nu_{-}} +c_1\left| x -1 \right|^{\nu_{+}},
\end{equation}
where $x= \rho/\rho_s$, constants $c_0$ and $c_1$ have different values to the left and right of the resonance, and the Mercier indices $\nu_{\pm}$ have values:
\begin{equation}
\nu_{\pm}=\frac{1}{2} \pm \sqrt{-D_M}. 
\end{equation}
This serves to define a generalised $\Delta^{\prime}$ 
\begin{equation}
\Delta^{\prime} = \left. \frac{c_1}{c_0} \right|_R + \left. \frac{c_1}{c_0} \right|_L, 
\end{equation}
where $R$ and $L$ denote locations immediately to the right and left of the resonance, respectively. This expression, 
obtained from the ideal MHD solution, must be matched to the analogous quantity arising from the 
inner resonant layer solution, to obtain the tearing mode dispersion relation.

Using the results in \cite{GGJ} and \cite{C2H3M}\footnote{A factor $1/f^2$ was missed from the final term of eqn.(B3) of 
\cite{C2H3M} due to a typographical error, and this is correctly included here.} 
we find that, at the tearing mode resonance,  $D$ of eqn.(\ref{eq:D}) should be compared to 
$-\frac{As^2}{\rho^2} \left( \frac{1}{4} + D_M \right)$, where $s=\rho q^{\prime}/q$ is the magnetic shear,
\begin{align} 
     D_M & = -\frac{1}{4} + E + F + H \nonumber \\
     & = -\frac{1}{4} + \frac{q}{q^{\prime}} \frac{\mu_0 p^{\prime}}{B_0^2 f^2} \left\langle \frac{R^2}{R_0^2} \frac{1}{|\nabla \rho|^2} \right\rangle 
             -  \frac{q^2}{q^{\prime 2}} \left( \frac{\mu_0 p^{\prime}}{B_0^2 f^2} \right)^2 \left\langle \frac{R^2}{R_0^2} \frac{1}{|\nabla \rho|^2} \right\rangle^2 \nonumber \\
         & - \left( \frac{\mu_0 p^{\prime}}{B_0^2 f^2} \right)  \frac{1}{q^{\prime 2}} \left( \frac{\rho}{R_0 f} \right)^2 \times 
 \left\langle \frac{\partial}{\partial \rho} \left( \frac{R^2}{R_0^2} \right) - \frac{R^2}{R_0^2} \frac{\rho}{f}\frac{d}{d\rho} \left(\frac{f}{\rho}\right) \right\rangle 
         \left\langle \frac{B^2 R^2}{B_0^2 R_0^2 |\nabla \rho|^2} \right\rangle \nonumber \\
         & + \left( \frac{\mu_0 p^{\prime}}{B_0^2 f^2 q^{\prime}} \right)^2 \left( \frac{\rho}{R_0 f} \right)^2  
         \left\langle \frac{R^4}{R_0^4 |\nabla \rho|^2} \right\rangle 
         \left\langle \frac{B^2 R^2}{B_0^2 R_0^2 |\nabla \rho|^2} \right\rangle \label{eq:DM},
\end{align}
and the quantities $E,~F $ and $H$ are defined in \cite{GGJ}. %Ref.[\cite{13}].  
(In a later paper, \cite{GGJ76} showed that for a large aspect ratio circular cross-section plasma:
\begin{equation}
 E+F+H = \frac{2 \rho \mu_0 p^{\prime}}{B_0^2} \frac{q^2-1}{s^2}
\end{equation}
where the important factor $q^2-1$ removes, for $q>1$, the possibility of the instability predicted by \cite{Suy} in a 
straight cylinder.)  Thus we can write:
\begin{equation}
 \frac{1}{4} + D_M \propto \frac{\rho \mu_0 p^{\prime}}{B_0^2} \frac{\kappa_{\rm eff}}{s^2},
\end{equation}
with the `effective' curvature , $\kappa_{\rm eff}$, deduced from eqn.(\ref{eq:DM}).
However, \cite{HC}, perhaps seeking a $D$ consistent with this argument, assumed $\kappa_{\rm eff}$ was the surface-averaged normal curvature, $\kappa_n$, 
and, furthermore, that $\kappa_n \propto V^{\prime \prime} = \frac{d \langle J \rangle}{d\rho}$, where $\langle J \rangle = \frac{\langle R^2 \rangle q}{R_0 B_0 g}$,  
to obtain the following result for $D$:
\begin{equation}
 D_{HC} \propto \frac{\rho \mu_0 p^{\prime}}{B_0^2 s^2} \frac{d \langle J \rangle}{d\rho} \propto \frac{\rho \mu_0 p^{\prime}}{B_0^2 s^2} \frac{1}{R_0} \frac{q}{g} \left( \frac{d \langle R^2 \rangle }{d \rho} 
 + \langle R^2 \rangle \left( \frac{q^{\prime}}{q} - \frac{g^{\prime}}{g} \right) \right). \label{eq:DCH}
\end{equation}
However, at low $\beta$ and with $B_{\phi} \simeq B$ (e.g. at large aspect ratio),  
\begin{equation}
 \kappa_n \propto V^{\prime\prime}-\frac{\langle R^2 \rangle}{R_0 B_0} \frac{q^{\prime}}{g}
\end{equation} \cite{CHH}, so that their argument should have implied
\begin{equation}
D_{HC} \rightarrow D \propto \frac{\rho \mu_0 p^{\prime}}{B_0^2 s^2} \frac{1}{R_0} \frac{q}{g} \left( \frac{d \langle R^2 \rangle }{d \rho} 
 - \langle R^2 \rangle \frac{g^{\prime}}{g} \right). \label{eq:DCHcorr}
\end{equation}
Equation~(\ref{eq:DCHcorr}) is indeed consistent with our expression for $D$ in eqn.(\ref{eq:D}), and also with the work of \cite{NCH} in
the special case $g^{\prime}=0$ that they considered. Equations~(\ref{eq:D},\ref{eq:DCHcorr}) are not, however, consistent with 
$D =-\frac{As^2}{\rho^2} \left( \frac{1}{4} + D_M \right)$, since $\kappa_{\rm eff} \ne \kappa_{n}$. 
\footnote{A more detailed discussion of the relation of $D_M$ to $\kappa_n$ is given by \cite{JG}.} 
We should not expect
$D$ to be exactly equal to $-\frac{As^2}{\rho^2} \left( \frac{1}{4} + D_M \right)$, 
because the ideal instability investigated by Mercier, 
and later by \cite{GandJ} using Hamada co-ordinates, is a mode with a range of coupled poloidal harmonics, 
whereas $\psi$ of the  envisaged tearing mode, has an isolated single poloidal harmonic. 
    
It would be inconsistent with the `single poloidal harmonic' assumption to simply replace $D(\rho)$ by the value corresponding to $D_M$ in eqn.(\ref{eq:1Dtear}); 
although the use of $D_M$ would capture the poloidal mode coupling effects close to the singular surface that 
can have a profound effect on the Mercier indices, which in turn influence the value of the generalised $\Delta^\prime$ stability parameter \cite{GGJ}.    
    
\section{ Conclusions}
    
Within the foregoing sections we have assumed that the perturbed poloidal flux function, $\psi(\rho,\theta)$,  contained only one poloidal harmonic, $e^{i m\theta}$. 
However our solution for the  variable $z$, eqn.(\ref{eq:dzdt}), contains a full spectrum of poloidal harmonics. Under these assumptions we have extended the validity of 
the tearing equation proposed by \cite{HC} 
to axisymmetric equilibria of arbitrary aspect ratio and arbitrary $\beta$. In doing so we have only made use of the approximation, $\epsilon^2/q_s^2 \ll1$. 
This would certainly rule out the use of the resulting 1-D equation for studying internal kink type disruptions in tokamaks 
(where the $m=n=1$ harmonic plays a crucial role), but should prove to be an accurate approximation for modes which are resonant in the 
pedestal region of a tokamak in H-mode.
An unexpected result of this calculation has been the appearance of a new set of terms arising from the effect of the integration constant $K$ 
(denoted by $\hat{A}, ~\hat{B}, ~\hat{C}$ and $\hat{D} $). However, it appears unlikely that such terms will play a significant role in 
determining tearing stability since they are normally negligibly small, except perhaps in very strongly shaped cross-sections or, 
e.g., in the vicinity of a separatrix boundary. For simplicity we ignored these extra terms in Eq.(\ref{eq:1Dtear_gen}).
    
It is also clear from the foregoing derivation of a 1-D equation that the pressure gradient term, 
$D(\rho)$ of eqn.(\ref{eq:1Dtear}), differs from the quantity $-\frac{As^2}{\rho^2} \left( \frac{1}{4} +D_M\right)$ that would be expected in general tearing mode theory, 
as the singular surface is approached. The difference arises because the derivation of 
eqn.(\ref{eq:1Dtear}) is based on a single poloidal harmonic assumption, 
whereas retention of the coupled poloidal harmonics is required to capture the true value in the limit as $\rho \to \rho_s$.
The approach outlined in \cite{FITZPATRICK_NF1993} retains seven coupled poloidal harmonics, but its restrictions to weak shaping and low $\beta$ 
severely impede its application to the pedestal. The single poloidal harmonic approach outlined in this paper accommodates strong
shaping and $\beta$ effects, but neglects poloidal mode coupling that is needed to describe $D_M$ at the resonance and that may be important more globally.
Nevertheless, for $\Delta^{\prime}$ calculations at the foot of the pedestal where $s^2$ becomes large near a separatrix  boundary, both the exact 
Mercier indices and the approximate (1-D) ones return to similar, low $\beta$, values (of 0 and 1),
and the 1-D approximation may give a good indication of tearing instability in a rather simple manner.

Numerical investigations of H-mode equilibria are presently underway.  

This work has been carried out under the RCUK Energy Programme [grant number EP/P012450/1] with partial funding through the framework of the EUROfusion Consortium  
from the Euratom research and training programme 2014-2018 under grant agreement No 633053. 
To obtain further information on the data and models underlying this paper please contact PublicationsManager@ukaea.uk. 
The views and opinions expressed herein do not necessarily reflect those of the European Commission.
The authors are grateful to Chris Ham for helpful discussions.  

\begin{appendices}
\section{}\label{ app A}
    
We can generate unique expressions for the coefficients $B,~C$ and $D$, by  exploiting the fact that all toroidal mode number dependencies 
in the 1-D tearing eqn.(\ref{eq:1Dtear}) can be expressed as powers, up to quadratic, of $\frac{m}{m-nq}$.
       
First, we collect all three terms in eqn.(\ref{eq:1Dtear_gen}), that 
include parts proportional to $\left(m/(m-nq)\right)^2$ and contribute to the coefficient $D(\rho)$ in eqn.(\ref{eq:1Dtear}), namely;
\begin{equation}
\frac{1}{ \lambda} \langle X \rangle + \rho \langle U^2 |\nabla \rho|^2 \rangle-\rho \frac{n}{m} q' \langle U | \nabla \rho |^2 \rangle
\end{equation}
Now replacing $n$ by the identity $(nq-m ~ +m)/q$ and using $s=\rho q^{\prime}/q$, this expression becomes
\begin{equation}
\frac{1}{ \lambda} \langle X \rangle + \rho \langle U^2 |\nabla \rho |^2 \rangle - s \langle U | \nabla \rho |^2 \rangle
 + \frac{(m-nq)}{m}s \langle U | \nabla \rho |^2 \rangle,
\end{equation} 
where the first three terms yield eqn.(\ref{eq:D}) for $D$  
and the last term now contributes to the expression for the coefficient $C$, 
rather than $D$. Three different terms from eqn.(\ref{eq:1Dtear_gen}) and the final term of eqn.(A2), contribute the term in eqn.(\ref{eq:1Dtear}) that is 
proportional to $m/(m-nq)$, with the following factor in the coefficient:
\begin{equation}
\frac{im}{ \lambda}\langle W \rangle+im\rho \langle U(T-T^*) |\nabla \rho |^2 \rangle 
- \frac{d}{d\rho} \left( \rho\langle U |\nabla \rho |^2\rangle \right) + s \langle U |\nabla \rho |^2 \rangle
\end{equation}
where the last term is the contribution from eqn.(A2) above. Using eqns.(\ref{eq:T}), (\ref{eq:U}) and (\ref{eq:W}) for $T$, $U$ and $W$, 
the expression in (A3) becomes:
\begin{equation}
-\frac{m}{n} \frac{d}{d\rho} \left[ \frac{R_0 g^\prime}{f}\right] 
- q\frac{d}{d\rho}\left[ \frac{R_0}{fg} \frac{\mu_0 p^\prime}{B_0^2} \langle \frac{R^2}{R_0^2}\rangle \right].
\end{equation}
Now, on replacing $m$ by the identity $m-nq+nq$, we obtain the following expression:
\begin{equation}
C =-q\frac{d}{d\rho}\left[ \frac{R_0 g^\prime}{f}+ \frac{R_0}{fg} \frac{\mu_0 p^\prime}{B_0^2} \left\langle \frac{R^2}{R_0^2}\right\rangle \right] 
-\frac{(m-nq)}{n} \frac{d}{d\rho} \left( \frac {R_0g^\prime}{f} \right),
\end{equation}
where the first two terms coincide with eqn.(\ref{eq:C}) for $C(\rho)$, and the third term contributes to the coefficient $B(\rho)$ and
exactly cancels the remaining $n$ dependence in $B$, leading to eqn.(\ref{eq:B}) for $B(\rho)$.

To demonstrate the second equality in eqn.(\ref{eq:B}) we consider cylindrical toroidal coordinates $R,Z,\phi$.  The Jacobian for the
transformation $(R,Z) \rightarrow (\rho,\theta)$ is:
\begin{equation}
 J = \frac{\rho R}{R_0} = \pd{R}{\theta} \pd{Z}{\rho} -\pd{R}{\rho} \pd{Z}{\theta} \label{eq:Jacobrhotheta}.
\end{equation}
We can obtain $\grad{\rho}$ and $\grad{\theta}$, using
\begin{align}
 \grad R & = \pd{R}{\rho} \grad{\rho} + \pd{R}{\theta} \grad \theta \nonumber \\
 \grad Z & = \pd{Z}{\rho} \grad \rho  + \pd{Z}{\theta} \grad \theta \label{eq:gradRZ}
\end{align} 
and deduce:
\begin{align}
 J^2 |\grad{\rho}|^2 & = \left(\pd{R}{\theta}\right)^2+ \left(\pd{Z}{\theta}\right)^2 \nonumber \\
 J^2 |\grad{\theta}|^2 & = \left(\pd{R}{\rho}\right)^2+ \left(\pd{Z}{\rho}\right)^2 \nonumber \\
 J^2 \grad{\rho}.\grad{\theta} & = - \left[ \pd{R}{\theta}\pd{R}{\rho}+ \pd{Z}{\theta}\pd{Z}{\rho} \right] \label{eq:J2}
\end{align}
Squaring and adding (\ref{eq:Jacobrhotheta}) and (\ref{eq:gradRZ}) using eqn.(\ref{eq:J2}) one finds:
\begin{equation}
 \frac{R^2}{R_0^2 |\grad{\rho}^2|}  + \rho^2 \frac{\grad{\theta}.\grad{\rho}}{|\grad{\rho}^2|}  = \rho^2 |\grad{\theta}|^2.
\end{equation}

\section{}\label{app B}
        
The 1-D tearing equation (\ref{eq:1Dtear}) is expressed in terms of equilibrium variables $\rho,~g$ and $f$. 
More familiar variables are  the equilibrium poloidal flux $\psi$ and $I(\psi)$ as used by \cite{HC}. %Ref.\cite{7}. 
These are related by eqns.(\ref{eq:I}-\ref{eq:psi}). 
In this appendix we give the form that eqn.(\ref{eq:1Dtear}) takes when expressed in these Hegna-Callen variables.
Of the four terms in eqn.(\ref{eq:1Dtear}) we find:
\begin{eqnarray}
    Term1&\rightarrow& \frac{I \rho}{q}\frac{d}{d \psi}\left[ \frac{q}{I} \langle |\nabla \psi|^2\rangle \frac{d{\tilde{A}}}{d \psi} \right],\\
    Term2 &\rightarrow& - m^2 \rho \langle | \nabla \theta |^2 \rangle \tilde{A},\\
    Term3 & \rightarrow& + \frac{ m \rho I }{(m-nq)} \frac{d}{d \psi} \left[ I^\prime(\psi)+\frac{\mu_0 p^\prime(\psi)}{I(\psi)}\langle R^2 \rangle \right] \tilde{A},\\
    Term4&\rightarrow& - \frac{ \rho I m^2 {\mu_0 p^{\prime}}}{(m-nq)^2} \left[ \frac{d}{d \psi} \left\langle \frac{R^2}{I} \right\rangle \right] \tilde{A}
\end{eqnarray}
where, as in the work of Hegna and Callen, the dependent variable $\tilde{A}$ is the, single poloidal harmonic, tearing mode eigenfunction, and
$^\prime$ denotes the radial derivative with respect to $\psi$.
Finally, on multiplying through by the factor $q/ \rho I$ we obtain the 1-D tearing equation in a rather simple form:
\small
    \begin{align}
    &\frac{d}{d \psi}\left[ \frac{q}{I} \langle |\nabla \psi|^2\rangle \frac{d{\tilde{A}}}{d \psi} \right] \nonumber \\
&    - \left\lbrace \frac{m^2 q}{I}\langle | \nabla \theta |^2 \rangle - \frac{ m q }{(m-nq)} \frac{d}{d \psi} 
    \left[ I^\prime+\frac{\mu_0 p^\prime}{I}\langle R^2 \rangle \right]+\frac{ m^2 q \mu_0 p^\prime}{(m-nq)^2}  \left[ \frac{d}{d \psi} 
    \left\langle \frac{R^2}{I} \right\rangle \right] \right\rbrace \tilde{A}=0, \label{eq:Btear}.
    \end{align}
\normalsize
\end{appendices}
	
\bibliographystyle{aip}
\bibliography{TearingJPP}

\end{document}